\documentclass{article}

\usepackage{array}
\usepackage{graphicx} 
\usepackage{soul} 
\usepackage{color}
\usepackage{url}
\usepackage{subcaption} 
\usepackage{amsmath} 
\usepackage{algorithm} 
\usepackage{algpseudocode}
\usepackage{multirow}
\usepackage{booktabs}
\usepackage{calc} 
\usepackage{calc} 
\usepackage{longtable}
\usepackage{amssymb} 
\usepackage{fvextra} 
\usepackage{xfp} 
\usepackage[super,sort&compress]{natbib}

\usepackage[a4paper,margin=3cm]{geometry}

\newenvironment{Shaded}{}{}

\newcommand{\NormalTok}[1]{#1}

\DefineVerbatimEnvironment{Highlighting}{Verbatim}{commandchars=\\\{\}}

\title{A benchmark suite of intracellular Boolean model variants and multiscale simulations for computational biology}

\author{
  Marco Masera\textsuperscript{2},
  Riccardo Smeriglio\textsuperscript{1},
  Roberta Bardini\textsuperscript{1},\\
  Alessandro Savino\textsuperscript{1},
  Stefano Di Carlo\textsuperscript{1}
  \\[0.5em]
  \small \textsuperscript{1}Department of Control and Computer Engineering, Politecnico di Torino, Turin, Italy \\
  \small \textsuperscript{2}Department of Life Sciences and Systems Biology, University of Turin, Turin, Italy \\
  \small Correspondence: Roberta Bardini, Riccardo Smeriglio (roberta.bardini@polito.it, riccardo.smeriglio@polito.it)
}

\date{March 2026}

\begin{document}

\maketitle

\begin{abstract}
We present PhysiBench, an open resource for developing and
evaluating computational methods in systems biology including a benchmark suite of
612 executable intracellular Boolean regulatory network variants and a
dataset of 120,000 time-resolved multiscale stochastic simulations. The
benchmark models are derived from seven published Boolean networks
spanning cell-cycle control, developmental patterning, cancer signaling,
immune response, and cell-fate decisions, and are executable in the
PhysiBoSS/PhysiCell multiscale simulation framework. Model variants are
generated through mutation-based model construction, online behavioral
filtering, and offline sensitivity evaluation. The simulation dataset is
produced from 60 selected models under systematically sampled stimulation
protocols and fixed model-level initial configurations. Each trajectory is
linked to its model identifier, input-parameter file, stochastic seed, and
cell-level output file. PhysiBench supports direct simulation, surrogate
modeling, data-driven inference, simulation-based optimization, and
comparative benchmarking. Technical validation includes file-integrity and
executability checks, graph-based structural diversity analyses, and
behavioral heterogeneity assessment from multiscale simulation outputs.
\end{abstract}

\section{Background \& Summary}\label{background-summary}

Computational systems biology increasingly relies on simulation-based
workflows to study, predict, and control complex biological processes.
Two methodological areas are especially dependent on reusable simulation
resources. The first is optimization via simulation, where search
algorithms repeatedly simulate a biological model to identify parameters,
perturbations, or stimulation protocols that produce a desired
behavior. Practical examples include calibrating uncertain model
parameters against observed dynamics or searching for stimulation and
treatment schedules that drive a simulated cell population toward a
target outcome \cite{ponce2022optimizing,castrignano2024,giannantoni2022}. The second is surrogate modeling, where statistical or
machine-learning emulators are trained to approximate computationally
expensive mechanistic simulations and predict their outputs at reduced
cost. Practical examples include training fast predictors of population
growth, spatial organization, or cell-fate trajectories from simulation
inputs, so that downstream optimization or uncertainty analysis can be
performed without repeatedly running full costly simulations \cite{abrate2024,norton2026}. Although these
fields address different tasks, they share a common need: controlled
biological simulation benchmarks that are executable, reproducible,
diverse, and paired with well-defined input spaces \cite{bardini2024computational,cascarano2023}.

Comparable benchmark resources have already accelerated method
development in other scientific domains. Datasets such as PDEBench
\cite{takamoto_pdebench_2022,pdebench_repo,darus_pdebench_dataset},
WeatherBench \cite{rasp_weatherbench_2020,pangeo_weatherbench_repo,google_weatherbench2_repo},
and The Well \cite{ohana2024well,polymathicai_the_well_repo} provide
standardized simulation data, controlled tasks, and reusable evaluation
settings for physics, climate modeling, and scientific machine learning.
Systems biology lacks an equivalent resource for the development of
simulation-based methods. Existing computational biology resources play
important complementary roles: BioModels, BioSimulators, and BioSimulations
\cite{maliksheriff2020,vaginay2022,shaikh2022},
PhysiBoSS-Models \cite{noel2025}, Boolean model repositories
\cite{pastva_bbm_biorxiv_2023,biodivine_boolean_models_repo}, and PhysiCell datasets
\cite{rocha_physicell_parameter_sensitivity_2025}
support model sharing, simulator discovery, reproducible execution, or
specific simulation studies. What they do not provide is a controlled
benchmark built around families of related executable biological models,
embedded in a common multiscale framework, and paired with large numbers
of systematically generated stochastic trajectories.

This gap limits both optimization and surrogate-modeling research in
systems biology. Researchers developing optimization via simulation
methods need executable biological models that can be repeatedly run
under controlled, comparable conditions. Researchers developing
surrogate models need large, paired input-output datasets with enough
variation to test whether emulators can learn nonlinear, stochastic,
multiscale dynamics. In the absence of such resources, methods are often
evaluated on ad hoc examples, making it difficult to compare algorithms,
assess robustness, or reproduce benchmark conditions. A further obstacle
is that building multiscale hybrid models by hand is resource-intensive,
requiring manual curation, domain expertise, and harmonization across
modeling formalisms \cite{ruscone2025neko}; this motivates pipelines that
instead generate diverse regulatory variants automatically by
systematically evolving a baseline logic \cite{ouellet2024}.

\begin{figure}[htbp]
    \centering
    \includegraphics[width=0.99\linewidth]{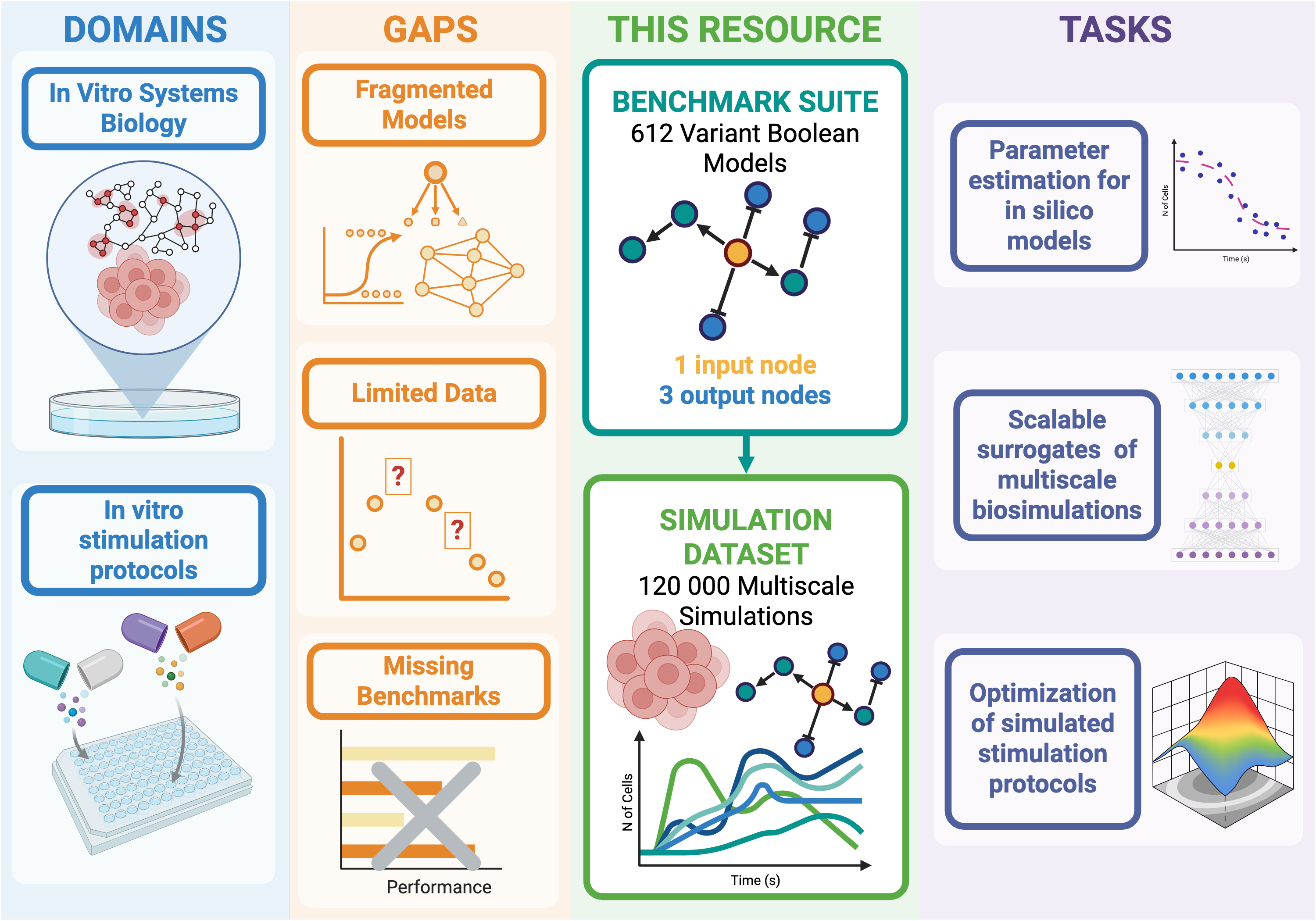}
    \caption{Overview of the resource, including a benchmark suite of 612 variant intracellular Boolean models embedded in a shared multiscale framework, for use by methods that generate dynamic data on demand, and a dataset of 120{,}000 precomputed multiscale simulations under controlled inputs, for use by methods that operate on time-resolved trajectories. Together, they provide inputs and reference trajectories for surrogate modeling, parameter estimation, simulation-based optimization, and benchmarking of controlled stimulation-protocol workflows.}
    \label{fig:overview}
\end{figure}

Here we present PhysiBench, a controlled in silico benchmark resource
designed to address this gap. The name reflects the resource's
foundation in the PhysiBoSS/PhysiCell ecosystem
\cite{stoll2017maboss,letort2019,calzone2022}, where intracellular
Boolean regulation is coupled to cell-level and spatial multiscale
simulation. PhysiBench consists of a
benchmark suite of 612 executable intracellular Boolean regulatory models, and a dataset of 120,000 time-resolved stochastic
multiscale simulations generated from 60 selected models
(Figure~\ref{fig:overview}). The model
suite provides a controlled family of related regulatory models
executable in a shared framework, while the simulation dataset provides
paired input-output trajectories generated under systematically varied
initial conditions and stimulation protocols.

The two components support complementary uses: the executable model
suite enables methods that require simulations on demand, whereas the
precomputed trajectories support surrogate modeling, emulator
evaluation, data-driven inference, and comparative benchmarking. Shared
identifiers and explicit input-parameter files link models, simulation
inputs, and outputs, allowing the resource to be used either as
executable benchmark problems or as a structured dataset for supervised
learning.

PhysiBench is biologically motivated but methodologically oriented. The
Boolean regulatory models are derived from published networks spanning
cell-cycle control, developmental patterning, cancer signaling, immune
response, and cell-fate decisions, but the released variants and
simulations are controlled in silico benchmarks rather than calibrated
representations of specific biological systems. This design provides
realistic biological structure and multiscale stochastic dynamics
\cite{bardini2017,alsalloum2024} while
preserving the reproducibility, scale, and controlled variation needed
for systematic method development in computational systems biology.

\section{Methods}\label{methods}

PhysiBench contains two linked outputs: a benchmark suite
of 612 executable intracellular Boolean regulatory network variants and
a dataset of 120,000 time-resolved multiscale simulation trajectories
generated from 60 selected models. The Methods are organized around
these two outputs: construction of the benchmark model suite and
generation of the trajectory dataset.

Throughout this section, a simulation context refers to the non-model
inputs required to define one PhysiBoSS run: the initial cell-population
configuration and the external stimulation protocol. The stimulation
protocol specifies the stimulation duration, stimulation period, and
four spatial boundary values defining the external signal field over the
two-dimensional simulation domain. Given a model, a context, and a stochastic seed, the corresponding simulation run is fully specified. The components included in the simulation context, and the run-specific metadata recorded separately, are summarized in Table~\ref{tab:simulation-context-components}.

\begin{longtable}[]{@{}
  >{\raggedright\arraybackslash}p{(\columnwidth - 4\tabcolsep) * \real{0.2189}}
  >{\raggedright\arraybackslash}p{(\columnwidth - 4\tabcolsep) * \real{0.1953}}
  >{\raggedright\arraybackslash}p{(\columnwidth - 4\tabcolsep) * \real{0.5858}}@{}}
\caption{Components of the simulation context and run-specific metadata.}
\label{tab:simulation-context-components}\\
\toprule\noalign{}
\begin{minipage}[b]{\linewidth}\raggedright
Component
\end{minipage} & \begin{minipage}[b]{\linewidth}\raggedright
Included in a simulation context?
\end{minipage} & \begin{minipage}[b]{\linewidth}\raggedright
Description
\end{minipage} \\
\midrule\noalign{}
\endhead
\bottomrule\noalign{}
\endlastfoot
Initial cell-population configuration & Yes & Defines the starting
spatial arrangement, density, and related initial-condition
parameters. \\
Stimulation duration & Yes & Duration of the external signal applied
through the model input node. \\
Stimulation period & Yes & Periodicity of the external stimulation
signal. \\
Spatial boundary values & Yes & Four boundary values defining the
external signal field over the two-dimensional simulation domain. \\
Model identifier & No & Specifies which candidate or selected model is
evaluated. \\
Stochastic seed & No & Recorded separately to reproduce each stochastic
simulation run. \\
\end{longtable}

The context schema and context files used at each workflow stage are
listed in the GitHub repository (see Section \ref{codeavail}).

\subsection{Construction of the benchmark model
suite}\label{construction-of-the-benchmark-model-suite}

\begin{figure}[htbp]
    \centering
    \includegraphics[width=\linewidth]{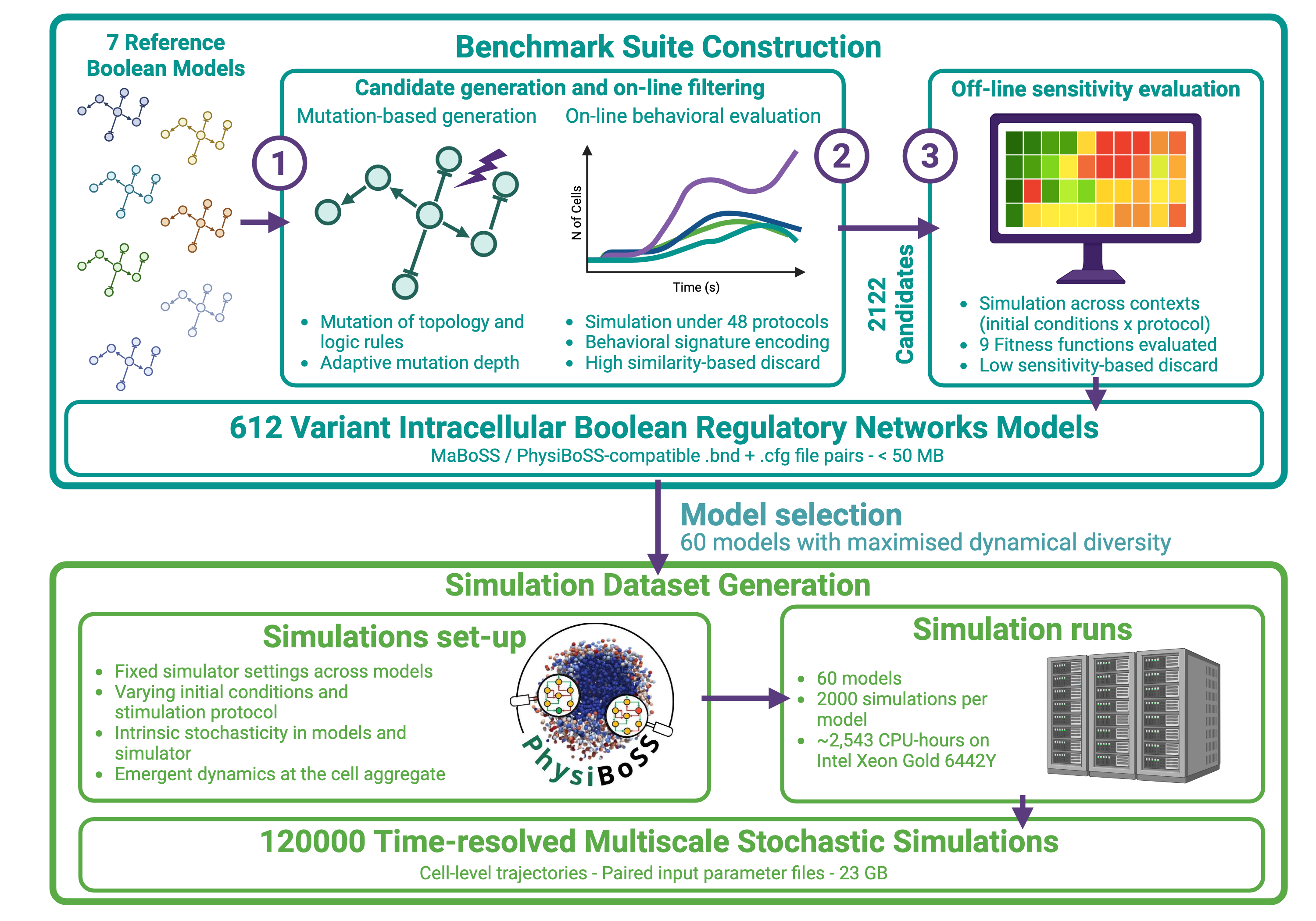}
    \caption{Overview of the pipeline. Starting from seven reference intracellular Boolean regulatory network models, benchmark suite construction proceeds through three stages: (1) mutation-based generation of candidate models, (2) online behavioral evaluation, and (3) offline sensitivity evaluation. This produces a benchmark suite of 612 variant intracellular Boolean regulatory network models. A subsequent model selection step retains 60 models to increase coverage of diverse dynamical regimes for simulation dataset generation. In the second phase, simulations are executed under fixed simulator settings and controlled variations in initial conditions and stimulation protocols, producing 120{,}000 time-resolved multiscale stochastic simulations.}
    \label{fig:pipeline}
\end{figure}

The benchmark-suite workflow starts from seven published source models,
assigns each model a common PhysiBoSS/PhysiCell interface, generates
stochastic model variants, and applies two filtering steps to retain
executable, nonredundant, and input-responsive variants (Figure~\ref{fig:pipeline}).

\subsubsection{Source Boolean models and common simulation
interface}\label{source-boolean-models-and-common-simulation-interface}

Seven published PhysiBoSS-compatible Boolean regulatory network models
are used as starting points for benchmark construction. They are
selected because they are open source, executable in the
PhysiBoSS/MaBoSS ecosystem, documented in previous studies, and cover
distinct biological processes and network architectures. The source
models are summarized in Table~\ref{tab:source-boolean-models}.

\begin{longtable}[]{@{}
  >{\raggedright\arraybackslash}p{(\columnwidth - 4\tabcolsep) * \real{0.3577}}
  >{\raggedright\arraybackslash}p{(\columnwidth - 4\tabcolsep) * \real{0.5691}}
  >{\raggedright\arraybackslash}p{(\columnwidth - 4\tabcolsep) * \real{0.0732}}@{}}

\caption{Published Boolean regulatory network models used as source models for benchmark construction.}
\label{tab:source-boolean-models}\\

\toprule\noalign{}

\begin{minipage}[b]{\linewidth}\raggedright
Source model
\end{minipage} & \begin{minipage}[b]{\linewidth}\raggedright
Short description
\end{minipage} & \begin{minipage}[b]{\linewidth}\raggedright
Reference
\end{minipage} \\
\midrule\noalign{}
\endhead
\bottomrule\noalign{}
\endlastfoot
Mammalian cell-cycle restriction-point model & Mammalian cell-cycle
restriction-point control. & \cite{faure2006,faure_ginsim_mammalian_cell_cycle} \\
\emph{Drosophila} segment-polarity model & Developmental patterning in
the \emph{Drosophila} segment-polarity network. & \cite{albert2003,stoll2010,maboss_drosophila_patterning} \\
EGF/TNF signaling model & Signaling responses to EGF and TNF
stimulation. & \cite{chaouiya2013,biomodels_biomd0000000562} \\
Gastric cancer signaling model & Regulatory mechanisms associated with
gastric cancer signaling. & \cite{flobak2015,physiboss_gastric_cancer} \\
Macrophage activation model & Immune-cell activation and macrophage
response states. & \cite{niarakis2024,physiboss_covid_boolean_network} \\
Prostate cancer signaling model & Regulatory mechanisms associated with
prostate cancer signaling. & \cite{montagud2022,physiboss_prostate_cancer} \\
TNF cell-fate model & TNF-induced cell-fate decisions. & \cite{calzone2010,physiboss_tnf_cell_fate} \\
\end{longtable}

Before variant generation, each reference model is prepared for use in
a common multiscale simulation workflow. This preparation focused on two
requirements. First, each model is checked for syntactic validity and
provided with compatible MaBoSS \texttt{.bnd} and \texttt{.cfg} files.
Second, each model is assigned the same high-level simulation
interface, allowing all variants to be controlled and observed
consistently during PhysiBoSS/PhysiCell simulations. The prepared model
files and interface assignments are released and
documented in the repository (See Section \ref{codeavail}).

The common interface consists of one intracellular input node and three
intracellular output nodes. The input node acts as the single
controllable signal through which the external stimulation protocol is
passed from the PhysiCell environment to the intracellular model. The
three output nodes are read by the multiscale simulator and used to
connect intracellular regulatory dynamics to cell-level behavior,
including survival, proliferation, and death.

This interface was introduced to standardize benchmarks. It should not
be interpreted as a biological claim that the selected nodes are
validated receptors, phenotypic markers, or therapeutic readouts for the
corresponding source systems.

\subsubsection{Generation of model
variants}\label{generation-of-model-variants}

Model variants are generated by applying stochastic modifications to
the seven prepared reference models. The goal was to create families of
related but distinct intracellular regulatory networks that could all be
executed through the same PhysiBoSS/PhysiCell interface.

For each candidate variant, one reference model is copied and modified
by a sequence of random mutation operations. These operations changed
the regulatory logic of the model while preserving the common simulation
interface. The single input node and the three output nodes assigned to
each reference model were protected from mutation, removal, or
reassignment. This ensured that every accepted variant could be
controlled by the same external stimulation protocol and observed
through the same intracellular outputs during multiscale simulation. The released mutation operators are summarized in Table~\ref{tab:mutation-operators}.

\begin{longtable}[]{@{}
  >{\raggedright\arraybackslash}p{(\columnwidth - 4\tabcolsep) * \real{0.1330}}
  >{\raggedright\arraybackslash}p{(\columnwidth - 4\tabcolsep) * \real{0.7340}}
  >{\raggedright\arraybackslash}p{(\columnwidth - 4\tabcolsep) * \real{0.1330}}@{}}

\caption{Mutation operators used to generate Boolean model variants.}
\label{tab:mutation-operators}\\

\toprule\noalign{}
\begin{minipage}[b]{\linewidth}\raggedright
Mutation operator
\end{minipage} & \begin{minipage}[b]{\linewidth}\raggedright
Effect on the model
\end{minipage} & \begin{minipage}[b]{\linewidth}\raggedright
Used in released generation
\end{minipage} \\
\midrule\noalign{}
\endhead
\bottomrule\noalign{}
\endlastfoot
\texttt{switch\ nodes\ logic} & Selects two nodes with defined logic and
swaps their logical expressions. & Yes \\
\texttt{replace\ logical\ operator} & Selects a binary logical operator
within a node rule and replaces it with another operator. & Yes \\
\texttt{replace\ node\ inside\ logic} & Selects a node appearing inside
a logical expression and replaces it with another node. & Yes \\
\texttt{negate\ subexpression} & Selects a subexpression within a
logical rule and adds or removes a logical negation. & Yes \\
\texttt{add\ input\ to\ logic} & Adds an additional input node to the
logical rule of a selected node by combining it with the existing
expression through a randomly chosen operator. & Yes \\
\texttt{add\ new\ node} & Creates a new regulatory node with a random
initial value and random logical rule, then connects it to existing
logic. & Yes \\
\texttt{randomize\ node\ logic} & Replaces the logical rule of a
selected node with a newly generated random expression built from
existing nodes. & Yes \\
\texttt{randomize\ parameter} & Randomizes one model parameter by
assigning it a value sampled uniformly between 0 and 1. & No \\
\end{longtable}

Mutation settings were chosen to favor local changes in regulatory logic
while preventing uncontrolled network growth. Operators that add new
nodes were assigned a low probability, and each mutation lineage was
limited to at most 45 added nodes. This limit was introduced as a
computational safeguard rather than a biological assumption: it allowed
substantial expansion of the reference networks while preventing
unbounded network growth that would reduce tractability, comparability
across variants, and traceability to the source models. After mutation,
each candidate was checked for compatibility with the MaBoSS and
PhysiBoSS execution formats. Candidates that failed these checks were
discarded.

The number of mutation operations applied to each candidate is adjusted
during generation to balance local exploration and broader search. When
generation produced acceptable variants, fewer mutations are applied to
subsequent candidates, encouraging exploration near successful models.
When candidates are repeatedly rejected, more mutations are applied,
increasing the chance of discovering distinct behaviors. The mutation
depth is bounded between 10 and 2,000 operations. The mutation code,
configuration file, and generated-model manifest are released and documented in the repository (See Section \ref{codeavail}).

\subsubsection{Online behavioral
filtering}\label{online-behavioral-filtering}

Each valid candidate model is evaluated immediately after generation to
reduce redundancy in the model suite. The candidate is simulated under
a fixed set of 48 stimulation protocols while holding the initial cell
configuration constant. For each protocol, the final six recorded time
points are retained, and the number of alive cells is extracted. These
values are concatenated across the 48 protocols to form a
population-level behavioral signature for the candidate. The exact 48
online-evaluation protocol definitions and
the sampled parameter space are provided in the GitHub repository (see Section \ref{codeavail}).
The candidate's signature is compared with the signatures of models
already accepted into the provisional suite.

Similarity is measured using Pearson correlation
\cite{benesty2009pearson} between behavioral signatures; candidates are
accepted only when their maximum correlation with the accepted pool was
below 0.85. Candidates whose behavior is too similar to previously
accepted models are discarded, whereas candidates with sufficiently
distinct population dynamics are retained. This online filtering step
produces 2,122 candidate variants from the seven reference architectures, as summarized in
Table~\ref{tab:online-filtering-candidates}.

\subsubsection{Offline sensitivity
filtering}\label{offline-sensitivity-filtering}

The online filtering step removes candidates that were behaviorally
redundant with previously accepted models. Offline sensitivity filtering
then removes candidates that are executable but weakly responsive to
controlled changes in simulation inputs.

For this step, each candidate model is simulated once across 215
released contexts. This produces, for each model, a set of outputs
spanning controlled variation in initial conditions and stimulation
protocols. The purpose of this step is not to evaluate biological
realism, but to retain models whose multiscale behavior changed
detectably across the sampled input space.

For every simulation, nine output summaries are computed. These
summaries measured total alive-cell count and spatial organization of
the alive-cell population with respect to circular and square target
regions, including shifted and distribution-weighted variants. 

For each model and each output summary, variability across the 215
contexts is quantified using the standard deviation and coefficient of
variation. A model is retained only when every output summary met both
thresholds: standard deviation greater than 30.0 and coefficient of
variation greater than 0.2. This rule removes models that are
executable but insensitive to the sampled input conditions. The output summaries, their code-level identifiers, 
the workflow scripts and configuration files are provided in the repository (See Section \ref{codeavail}). 

After offline filtering, 612 Boolean
regulatory network variants are retained as the released benchmark
suite.

The number of retained variants differs across reference families
because some source architectures generated more diverse multiscale
behaviors under the mutation and filtering procedure, as summarized in
Table~\ref{tab:online-filtering-candidates}.

\begin{longtable}[]{@{}
  >{\raggedright\arraybackslash}p{(\columnwidth - 4\tabcolsep) * \real{0.3793}}
  >{\raggedright\arraybackslash}p{(\columnwidth - 4\tabcolsep) * \real{0.2672}}
  >{\raggedright\arraybackslash}p{(\columnwidth - 4\tabcolsep) * \real{0.3534}}@{}}

\caption{Candidate variants retained after online behavioral filtering and offline sensitivity filtering for each source model family.}
\label{tab:online-filtering-candidates}\\

\toprule\noalign{}
\begin{minipage}[b]{\linewidth}\raggedright
Source model
\end{minipage} & \begin{minipage}[b]{\linewidth}\raggedright
Variants after online filtering
\end{minipage} & \begin{minipage}[b]{\linewidth}\raggedright
Variants retained after offline filtering
\end{minipage} \\
\midrule\noalign{}
\endhead
\bottomrule\noalign{}
\endlastfoot
Mammalian cell-cycle restriction-point model & 204 & 56 \\
\emph{Drosophila} segment-polarity model & 226 & 84 \\
EGF/TNF signaling model & 263 & 103 \\
Gastric cancer signaling model & 48 & 15 \\
Macrophage activation model & 45 & 15 \\
Prostate cancer signaling model & 51 & 18 \\
TNF cell-fate model & 1,285 & 321 \\
\textbf{Total} & \textbf{2,122} & \textbf{612} \\
\end{longtable}

The main hyperparameters used for benchmark-suite construction are
summarized in Table~\ref{tab:benchmark-construction-hyperparameters} to
support reproducibility. 
The full configuration files and execution environment are described in the GitHub repository (see Section \ref{codeavail}).

\begin{longtable}[]{@{}
  >{\raggedright\arraybackslash}p{(\columnwidth - 4\tabcolsep) * \real{0.3191}}
  >{\raggedright\arraybackslash}p{(\columnwidth - 4\tabcolsep) * \real{0.5532}}
  >{\raggedright\arraybackslash}p{(\columnwidth - 4\tabcolsep) * \real{0.1277}}@{}}

\caption{Main hyperparameters used for benchmark-suite construction.}
\label{tab:benchmark-construction-hyperparameters}\\

\toprule\noalign{}
\begin{minipage}[b]{\linewidth}\raggedright
Focus
\end{minipage} & \begin{minipage}[b]{\linewidth}\raggedright
Hyperparameter
\end{minipage} & \begin{minipage}[b]{\linewidth}\raggedright
Value
\end{minipage} \\
\midrule\noalign{}
\endhead
\bottomrule\noalign{}
\endlastfoot
Mutation-based generation & Evaluation protocols per behavioral
signature & 48 \\
Mutation-based generation & Final time steps retained per protocol &
6 \\
Mutation-based generation & Novelty threshold & 0.15 \\
Mutation-based generation & Mutation depth, initial and minimum & 10 \\
Mutation-based generation & Mutation depth, maximum & 2,000 \\
Mutation-based generation & Adaptive decay factor after candidate
acceptance & \(\times\) 0.75 \\
Mutation-based generation & Adaptive growth factor after candidate
rejection & \(\times\) 1.5 \\
Mutation operator distribution & \texttt{switch\ nodes\ logic} & 0.15 \\
Mutation operator distribution & \texttt{replace\ logical\ operator} &
0.26 \\
Mutation operator distribution & \texttt{replace\ node\ inside\ logic} &
0.228 \\
Mutation operator distribution & \texttt{negate\ subexpression} &
0.25 \\
Mutation operator distribution & \texttt{add\ input\ to\ logic} &
0.01 \\
Mutation operator distribution & \texttt{add\ new\ node} & 0.002 \\
Mutation operator distribution & \texttt{randomize\ node\ logic} &
0.1 \\
Mutation operator distribution & \texttt{randomize\ parameter} & 0
(disabled) \\
Network constraints & Maximum new nodes per lineage & 45 \\
Network constraints & Maximum network size after interface
standardization & 63 \\
Offline filtering & Simulation contexts per model & 215 \\
Offline filtering & Output summaries & 9 \\
Offline filtering & Minimum standard deviation & 30.0 \\
Offline filtering & Minimum coefficient of variation & 0.2 \\
Offline filtering & MaBoSS pseudorandom seed & 37 \\
Pipeline termination & Maximum candidates tested & 200,000 \\
Pipeline termination & Generated models before offline filtering &
2,122 \\
Pipeline termination & Models retained after offline filtering & 612 \\
\end{longtable}

\subsection{Generation of the multiscale trajectory
dataset}\label{generation-of-the-multiscale-trajectory-dataset}

The trajectory-dataset workflow starts from the 612-model benchmark
suite, selects 60 models to increase coverage of variable and extreme
simulation behaviors, and simulates each selected model under 2,000
controlled contexts.

\subsubsection{Selection of models for dataset
generation}\label{selection-of-models-for-dataset-generation}

A subset of 60 models was selected from the 612-model benchmark suite to
increase coverage of diverse and informative simulation behaviors while
keeping dataset generation computationally tractable.

Thirty models are selected for high variability in alive-cell-count
outputs across sampled contexts. Pairwise correlations \cite{benesty2009pearson} are used to
reduce redundancy: when two selected models produce highly correlated
response profiles, the model with lower variability is replaced by the
next eligible candidate. A second group of 30 models is selected for
strong outlier behavior, based on the frequency and magnitude of extreme
output responses across contexts. Outlier responses were identified
using the modified z-score, with values above 3.5 in absolute value
classified as outliers. Models selected by the first criterion were
excluded from the second criterion, ensuring that the two groups were
disjoint.

The two complementary selection groups used to choose the final 60 models
for simulation-dataset generation are summarized in
Table~\ref{tab:model-selection-groups}.

\begin{longtable}[]{@{}
  >{\raggedright\arraybackslash}p{(\columnwidth - 6\tabcolsep) * \real{0.1399}}
  >{\raggedright\arraybackslash}p{(\columnwidth - 6\tabcolsep) * \real{0.0546}}
  >{\raggedright\arraybackslash}p{(\columnwidth - 6\tabcolsep) * \real{0.3276}}
  >{\raggedright\arraybackslash}p{(\columnwidth - 6\tabcolsep) * \real{0.4778}}@{}}
  
\caption{Model-selection groups used to choose the 60 benchmark models for simulation-dataset generation.}
\label{tab:model-selection-groups}\\

\toprule\noalign{}
\begin{minipage}[b]{\linewidth}\raggedright
Selection group
\end{minipage} & \begin{minipage}[b]{\linewidth}\raggedright
Number of models
\end{minipage} & \begin{minipage}[b]{\linewidth}\raggedright
Selection criterion
\end{minipage} & \begin{minipage}[b]{\linewidth}\raggedright
Redundancy control
\end{minipage} \\
\midrule\noalign{}
\endhead
\bottomrule\noalign{}
\endlastfoot
High-variability models & 30 & Largest standard deviation in
alive-cell-count outputs across sampled contexts. & Pairwise
correlations were checked; when correlation exceeded 0.85, the
lower-variability model was replaced by the next eligible candidate. \\
Outlier-response models & 30 & Highest outlier-strength scores,
capturing both the frequency and severity of extreme responses. & Models
selected by the high-variability criterion were excluded so that the
groups remained disjoint. \\
\textbf{Total selected for dataset generation} & \textbf{60} &
Complementary coverage of broadly variable and rare or extreme
behaviors. & Selection groups were kept disjoint. \\
\end{longtable}

Together, these two criteria retained models with broadly variable
responses and rare or extreme responses. The selected model identifiers,
selection scores, and model-selection script are released with the
workflow (See Section \ref{codeavail}).

\subsubsection{Sampling of simulation
contexts}\label{sampling-of-simulation-contexts}

For each selected model, 2,000 contexts are generated using the shared
context definition introduced above. For each protocol parameter, ten
evenly spaced candidate values are generated over the configured range.
Stimulation duration is sampled between 5 and 200 simulation-time
units, stimulation period between 5 and 800 simulation-time units, and
each spatial boundary value between 0 and 10. The resulting grid is
filtered to retain only protocols in which the stimulation period was
greater than or equal to the stimulation duration. The remaining
protocols are randomly shuffled, and 2,000 contexts are retained for
each selected model. Contexts are sampled independently for each model,
so the exact set of stimulation protocols differs between models. The
sampled dataset-generation contexts and the sampling script are released
with the repository (See Section \ref{codeavail}).

The parameter ranges and sampling rules used to generate stimulation
protocols for simulation-dataset construction are summarized in
Table~\ref{tab:simulation-context-sampling}.

\begin{longtable}[]{@{}
  >{\raggedright\arraybackslash}p{(\columnwidth - 6\tabcolsep) * \real{0.2088}}
  >{\raggedright\arraybackslash}p{(\columnwidth - 6\tabcolsep) * \real{0.2021}}
  >{\raggedright\arraybackslash}p{(\columnwidth - 6\tabcolsep) * \real{0.1280}}
  >{\raggedright\arraybackslash}p{(\columnwidth - 6\tabcolsep) * \real{0.4611}}@{}}

\caption{Parameter ranges and sampling rules used to generate simulation contexts.}
\label{tab:simulation-context-sampling}\\

\toprule\noalign{}
\begin{minipage}[b]{\linewidth}\raggedright
Parameter or rule
\end{minipage} & \begin{minipage}[b]{\linewidth}\raggedright
Candidate values
\end{minipage} & \begin{minipage}[b]{\linewidth}\raggedright
Range or value
\end{minipage} & \begin{minipage}[b]{\linewidth}\raggedright
Constraint or sampling rule
\end{minipage} \\
\midrule\noalign{}
\endhead
\bottomrule\noalign{}
\endlastfoot
Stimulation duration & 10 evenly spaced values & 5--200 simulation-time
units & Stimulation period must be greater than or equal to stimulation
duration. \\
Stimulation period & 10 evenly spaced values & 5--800 simulation-time
units & Stimulation period must be greater than or equal to stimulation
duration. \\
\texttt{xmin} boundary value & 10 evenly spaced values & 0--10 & None
beyond grid sampling. \\
\texttt{xmax} boundary value & 10 evenly spaced values & 0--10 & None
beyond grid sampling. \\
\texttt{ymin} boundary value & 10 evenly spaced values & 0--10 & None
beyond grid sampling. \\
\texttt{ymax} boundary value & 10 evenly spaced values & 0--10 & None
beyond grid sampling. \\
Protocol grid & 1,000,000 combinations before filtering & Six parameters
with 10 candidate values each & Invalid protocols with stimulation
period shorter than stimulation duration were removed. \\
Retained contexts & 2,000 per selected model & 120,000 total simulations
across 60 models & Valid protocols were randomly shuffled and retained
independently for each model. \\
\end{longtable}

\subsubsection{Multiscale simulation
execution}\label{multiscale-simulation-execution}

All simulations are executed in the PhysiBoSS/PhysiCell framework
\cite{letort2019,calzone2022}. Each
cell contains an independent MaBoSS Boolean network instance that is
updated asynchronously during the simulation \cite{stoll2017maboss}. The intracellular input
node receives the external stimulation signal from the PhysiCell
environment, and the intracellular output nodes are propagated back to
the cell-based layer to modulate cell behavior. This coupling allows
intracellular regulatory dynamics, cell-level decisions, spatial
interactions, and environmental fields to evolve jointly.

For each of the 60 selected models, 2,000 independent simulations runs produce 120,000 time-resolved multiscale stochastic simulations.
Each simulation was assigned a unique stochastic seed recorded with its
input parameters. Simulator-level settings were held fixed for a given
Boolean model, including intracellular update settings, maximum
simulation time, cell mechanics parameters, microenvironmental diffusion
and decay parameters, boundary conditions, and PhysiBoSS coupling
settings. Therefore, variation in the released dataset arises from
controlled differences in initial conditions and stimulation protocols,
together with the intrinsic stochasticity of the multiscale simulator.

The final multiscale trajectory dataset is summarized in
Table~\ref{tab:simulation-dataset-summary}.

\begin{longtable}[]{@{}
  >{\raggedright\arraybackslash}p{(\columnwidth - 2\tabcolsep) * \real{0.3546}}
  >{\raggedright\arraybackslash}p{(\columnwidth - 2\tabcolsep) * \real{0.6454}}@{}}

\caption{Summary of the multiscale simulation dataset generated from the selected benchmark models.}
\label{tab:simulation-dataset-summary}\\

\toprule\noalign{}
\begin{minipage}[b]{\linewidth}\raggedright
Dataset quantity or setting
\end{minipage} & \begin{minipage}[b]{\linewidth}\raggedright
Value or description
\end{minipage} \\
\midrule\noalign{}
\endhead
\bottomrule\noalign{}
\endlastfoot
Selected models & 60 \\
Simulations per selected model & 2,000 \\
Total simulations & 120,000 \\
Stochastic seed & One recorded seed per simulation \\
Controlled variation & Initial conditions and stimulation protocols \\
Fixed model-specific settings & Intracellular update settings, maximum
simulation time, cell mechanics, microenvironmental diffusion and decay,
boundary conditions, and PhysiBoSS coupling settings \\
Recorded cell-level variables & x position, y position, z position, and
current phase \\
\end{longtable}

The fixed PhysiBoSS/PhysiCell configuration files used for simulation
execution are released with the workflow (See Section \ref{codeavail}).

Each simulation output records the cell population over time. For each
recorded time point, the released files include cell-level spatial
coordinates and cell-cycle phase information, allowing users to
reconstruct spatial trajectories and compute additional population-level
summaries. Each simulation is linked to an input-parameter file containing the
sampled stimulation protocol and to model-level initial-configuration
metadata. Together with the model identifier and stochastic seed recorded
in the manifest, these records provide the information required to
reproduce the run. The trajectory manifest links each simulation output
to its model identifier, sampled context, stochastic seed,
input-parameter file, and output file. 

\subsection{Resource organization and
reproducibility}\label{resource-organization-and-reproducibility}

The released repository provides the model files, context files,
simulation configurations, trajectory manifest, filtering scripts, and
workflow documentation needed to reproduce the benchmark-suite
construction and dataset-generation workflow (see Section \ref{codeavail}). This organization allows the resource to be used either as an
executable benchmark suite or as a precomputed trajectory dataset.

\section{Data Records}\label{data-records}

The PhysiBench resource as an open-access record including
the benchmark model suite and the multiscale trajectory dataset {[}Repository/DOI: add
record{]}. The benchmark suite contains the 612
executable intracellular Boolean regulatory network variants. The
trajectory dataset contains 120,000 time-resolved multiscale simulations
generated from 60 selected benchmark models. The two records can be
reused independently or linked through shared model identifiers.

The two linked components of the PhysiBench resource are summarized in
Table~\ref{tab:resource-components}.

\begin{longtable}[]{@{}
  >{\raggedright\arraybackslash}p{(\columnwidth - 6\tabcolsep) * \real{0.1551}}
  >{\raggedright\arraybackslash}p{(\columnwidth - 6\tabcolsep) * \real{0.6524}}
  >{\raggedright\arraybackslash}p{(\columnwidth - 6\tabcolsep) * \real{0.1016}}
  >{\raggedright\arraybackslash}p{(\columnwidth - 6\tabcolsep) * \real{0.0909}}@{}}

\caption{Main components of the released PhysiBench resource.}
\label{tab:resource-components}\\

\toprule\noalign{}
\begin{minipage}[b]{\linewidth}\raggedright
Component
\end{minipage} & \begin{minipage}[b]{\linewidth}\raggedright
Content
\end{minipage} & \begin{minipage}[b]{\linewidth}\raggedright
Number of records
\end{minipage} & \begin{minipage}[b]{\linewidth}\raggedright
Storage footprint
\end{minipage} \\
\midrule\noalign{}
\endhead
\bottomrule\noalign{}
\endlastfoot
Benchmark model suite & Variant intracellular Boolean regulatory network
models compatible with PhysiBoSS & 612 models & \textless{}50 MB \\
Multiscale trajectory dataset & Time-resolved multiscale simulation
outputs generated from 60 selected benchmark models under controlled
parameterizations & 120,000 simulations & 23 GB \\
\end{longtable}

\subsection{Benchmark model suite}\label{benchmark-model-suite}

The benchmark model suite contains 612 model variants organized by
reference family and model identifier. Each model is provided as a
MaBoSS-compatible pair of files: a \texttt{.bnd} file defining the
regulatory nodes, logical rules, and transition rates, and a
\texttt{.cfg} file specifying simulation settings. The minimum reusable
unit is one model variant, identified by the triplet
\texttt{\textless{}reference\_model,\ variant\_ID,\ model\_ID\textgreater{}}.

A manifest file, \texttt{variant\_models\_manifest.json}, indexes all
612 models and records the reference model, variant identifier, and
model identifier for each entry. This manifest allows users to
programmatically enumerate models, filter by reference family, and link
selected models to the simulation dataset.
The main fields of the benchmark-suite manifest are summarized in
Table~\ref{tab:benchmark-model-manifest-fields}.

\begin{longtable}[]{@{}
  >{\raggedright\arraybackslash}p{(\columnwidth - 2\tabcolsep) * \real{0.1954}}
  >{\raggedright\arraybackslash}p{(\columnwidth - 2\tabcolsep) * \real{0.8046}}@{}}

\caption{Main fields of the benchmark model-suite manifest.}
\label{tab:benchmark-model-manifest-fields}\\

\toprule\noalign{}
\begin{minipage}[b]{\linewidth}\raggedright
Field
\end{minipage} & \begin{minipage}[b]{\linewidth}\raggedright
Description
\end{minipage} \\
\midrule\noalign{}
\endhead
\bottomrule\noalign{}
\endlastfoot
\texttt{reference\_model} & Source reference family from which the
variant was generated. \\
\texttt{variant\_ID} & Variant identifier within the reference
family. \\
\texttt{model\_ID} & Model identifier used to link benchmark models and
simulation outputs. \\
\texttt{bnd\_file} & Path to the MaBoSS \texttt{.bnd} file. \\
\texttt{cfg\_file} & Path to the MaBoSS \texttt{.cfg} file. \\
\end{longtable}

\subsection{Multiscale trajectory
dataset}\label{multiscale-trajectory-dataset}

The simulation dataset contains 120,000 time-resolved multiscale
simulations generated from 60 selected benchmark models, with 2,000
simulations per selected model. The minimum reusable unit is one
simulation, identified by the quadruplet
\texttt{\textless{}reference\_model,\ variant\_ID,\ model\_ID,\ simulation\_ID\textgreater{}}.

A manifest file, \texttt{multiscale\_simulations\_manifest.json}, indexes all simulations and links each simulation to its parent model through the fields \texttt{reference\_model}, \texttt{variant\_ID}, \texttt{model\_ID}, and \texttt{simulation\_ID}. The corresponding input-parameter and cell-data files are organized under the same model-specific directory structure.

\subsection{Simulation-context metadata}\label{simulation-context-metadata}

Each PhysiBoSS simulation run is specified by a Boolean regulatory model, a stochastic seed, an initial cell-population configuration, and a stimulation protocol. The stimulation protocol is stored in the corresponding \texttt{input\_parameters} file, whereas the initial configuration is defined at model level in \texttt{initial\_positions.json} and is shared by simulations associated with the same \texttt{model\_ID}.

The released input-parameter files retain the field names \texttt{treatment\_duration} and \texttt{treatment\_period}; throughout the manuscript, these fields are described as stimulation duration and stimulation period. The field names follow the released configuration files.

The fields defining model-level initial configurations and
simulation-level stimulation protocols are summarized in
Table~\ref{tab:simulation-context-metadata-fields}.

\begin{longtable}[]{@{}
>{\raggedright\arraybackslash}p{(\columnwidth - 4\tabcolsep) * \real{0.2600}}
>{\raggedright\arraybackslash}p{(\columnwidth - 4\tabcolsep) * \real{0.2600}}
>{\raggedright\arraybackslash}p{(\columnwidth - 4\tabcolsep) * \real{0.4800}}@{}}

\caption{Fields used to define initial configurations and stimulation protocols in the simulation dataset.}
\label{tab:simulation-context-metadata-fields}\\

\toprule\noalign{}
\begin{minipage}[b]{\linewidth}\raggedright
Field
\end{minipage} & \begin{minipage}[b]{\linewidth}\raggedright
Source
\end{minipage} & \begin{minipage}[b]{\linewidth}\raggedright
Description
\end{minipage} \\
\midrule\noalign{}
\endhead
\bottomrule\noalign{}
\endlastfoot
\texttt{type} & \texttt{initial\_positions.json} &
Geometry of the initial cell population, such as circle or square. \\
\texttt{center} & \texttt{initial\_positions.json} &
Center coordinates of the initial cell population. \\
\texttt{density} & \texttt{initial\_positions.json} &
Fraction of candidate lattice positions occupied by cells. \\
\texttt{cell\_type} & \texttt{initial\_positions.json} &
Simulated cell type identifier. \\
\texttt{mode} & \texttt{initial\_positions.json} &
Initial-cell sampling mode, such as sparse or contour. \\
\texttt{length} & \texttt{initial\_positions.json} &
Radius for circular initializations or half-side length for square
initializations. \\
\texttt{treatment\_period} & \texttt{input\_parameters} &
Periodicity of the external stimulation signal, in simulation time units. \\
\texttt{treatment\_duration} & \texttt{input\_parameters} &
Duration of each stimulation interval, in simulation time units. \\
\texttt{TNF\_dirichlet\_xmin},
\texttt{TNF\_dirichlet\_xmax},
\texttt{TNF\_dirichlet\_ymin},
\texttt{TNF\_dirichlet\_ymax} & \texttt{input\_parameters} &
Boundary values defining the external stimulation field over the
two-dimensional simulation domain. \\
\end{longtable}

\subsection{Cell-data output files}\label{cell-data-output-files}

Time-resolved cellular outputs are stored in compressed
\texttt{cell\_data} files. Each file contains the saved time points for
one simulation. At each saved time point, the released records contain the
cell positions and the corresponding cell-cycle or death-state annotation,
which are sufficient to reconstruct spatial trajectories and compute
population-level summaries.

The core fields used in this manuscript are:

\begin{itemize}
\item \texttt{time}: simulation time, expressed in simulation time units;
\item \texttt{x\_positions}: x coordinates of the simulated cells;
\item \texttt{y\_positions}: y coordinates of the simulated cells;
\item \texttt{z\_positions}: z coordinates of the simulated cells;
\item \texttt{current\_phase}: PhysiCell cell-cycle or death-state phase
  associated with each cell.
\end{itemize}

Spatial coordinates follow the PhysiCell unit convention and are reported
in micrometers. Simulation times, saved intervals, stimulation periods, and
stimulation durations are reported in simulation time units. The
\texttt{current\_phase} field follows the PhysiCell integer phase encoding
used in the released simulations: live cells are encoded as 14, apoptotic
cells as 100, and necrotic states as 101, 102, or 103. Additional
PhysiCell-derived quantities, when present in exported simulation outputs,
follow the corresponding PhysiCell unit conventions.

\subsection{Directory organization}\label{directory-organization}

The benchmark suite and trajectory dataset are organized by shared model
identifiers. The expected directory organization is:

\begin{Shaded}
\begin{Highlighting}[]
\NormalTok{benchmark_suite/}
\NormalTok{|{-}{-} variant_models_manifest.json}
\NormalTok{|{-}{-} \textless{}reference_model\textgreater{}_\textless{}variant_ID\textgreater{}_\textless{}model_ID\textgreater{}/}
\NormalTok{|   |{-}{-} model.bnd}
\NormalTok{|   \textasciigrave{}{-}{-} model.cfg}
\NormalTok{\textasciigrave{}{-}{-} ...}
\end{Highlighting}
\end{Shaded}

\begin{Shaded}
\begin{Highlighting}[]
\NormalTok{simulation_dataset/}
\NormalTok{|{-}{-} multiscale_simulations_manifest.json}
\NormalTok{|{-}{-} initial_positions.json}
\NormalTok{|{-}{-} \textless{}reference_model\textgreater{}_\textless{}variant_ID\textgreater{}_\textless{}model_ID\textgreater{}/}
\NormalTok{|   \textasciigrave{}{-}{-} data/}
\NormalTok{|       |{-}{-} input_parameters/}
\NormalTok{|       |   |{-}{-} input_parameters_\textless{}simulation_ID\textgreater{}.json}
\NormalTok{|       |   \textasciigrave{}{-}{-} ...}
\NormalTok{|       \textasciigrave{}{-}{-} cell_data/}
\NormalTok{|           |{-}{-} cell_data_\textless{}simulation_ID\textgreater{}.json.gz}
\NormalTok{|           \textasciigrave{}{-}{-} ...}
\NormalTok{\textasciigrave{}{-}{-} ...}
\end{Highlighting}
\end{Shaded}

Time-resolved simulation outputs are stored as compressed JSON files
(\texttt{.json.gz}) and can be parsed using standard JSON libraries after
decompression. The \texttt{input\_parameters} files are JSON metadata
records associated with individual simulations. The manifests provide the
primary entry point for programmatic access, filtering, and linking
between benchmark models, initial configurations, simulation inputs,
stochastic seeds, and trajectory outputs.

\section{Technical Validation}\label{technical-validation}

The resource is validated through three complementary checks: file
integrity and executability, structural diversity of the benchmark
models, and behavioral heterogeneity of the multiscale outputs. These
checks assess whether the released files are complete, readable,
executable in the intended simulation framework, and diverse with
respect to the structural and behavioral metrics reported below.

\subsection{Resource integrity and
executability}\label{resource-integrity-and-executability}

For the benchmark suite, all 612 model directories are checked for the
expected \texttt{.bnd} and \texttt{.cfg} files. Each model is verified
to be syntactically valid, fully specified, and compatible with the
PhysiBoSS execution format. The benchmark manifest is cross-checked
against the directory structure to confirm that each listed model
identifier corresponds to an available model directory and that no
duplicate model identifiers are present.

For the simulation dataset,
\texttt{multiscale\_simulations\_manifest.json} is checked against the
released directory structure to verify the presence of 120,000
simulation records, corresponding to 2,000 simulations for each of the
60 selected benchmark models.
For each simulation record, the corresponding input-parameter file and
cell-data file are verified to be present and parsable with standard JSON
libraries, with decompression applied where required. Input-parameter
files are checked for the required stimulation-protocol fields, and the
manifest is checked for the metadata required to link each simulation to
its model identifier, initial configuration, stochastic seed, and
trajectory output.

Executability is assessed by running representative PhysiBoSS
simulations from the released model and configuration files without
model-specific adaptation. These checks confirm that the benchmark suite
can be used as an executable collection of intracellular Boolean
regulatory network models and that the simulation dataset is
consistently indexed, readable, and linked to the corresponding model
and input records.

\subsection{Structural diversity of the benchmark
suite}\label{structural-diversity-of-the-benchmark-suite}

Structural diversity is assessed using graph-based distance measures to
verify that mutation-based generation produced distinct regulatory
structures rather than minor syntactic variants. Boolean update rules
are converted into graph representations, and each model is mapped to
a directed signed graph whose edges encode regulatory dependencies and
polarity.

\begin{figure}[htbp]
\centering
\includegraphics[width=1\textwidth]{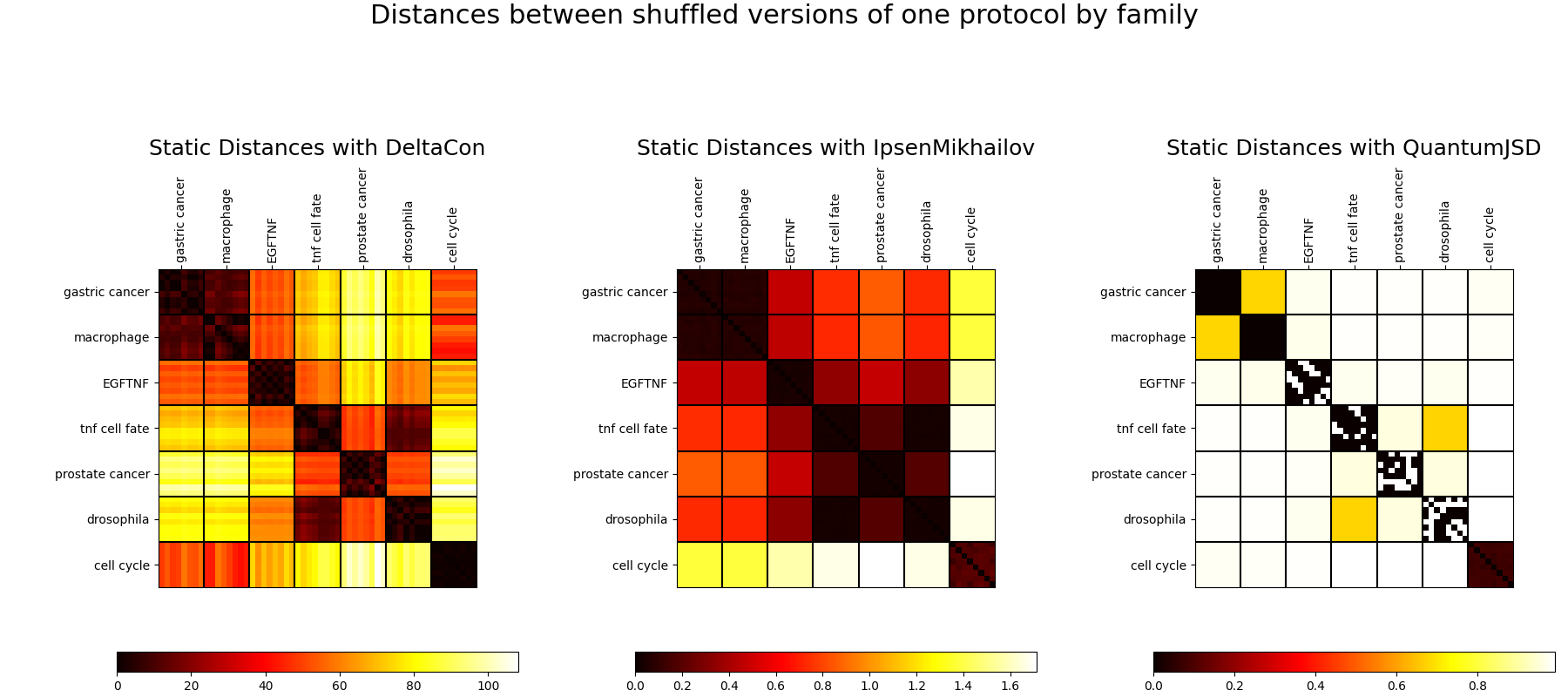}
    \caption{Pairwise structural distances computed on the Control Set used to validate the graph-based similarity metrics. Panels show heatmaps of (a) DeltaCon, (b) Ipsen-Mikhailov distance, and (c) Quantum Jensen-Shannon divergence computed on one original reference model and a set of shuffled versions of it, for each reference model. Darker blocks along the diagonal correspond to shuffled instances derived from the same reference model and indicate lower distances than those observed between different source models.}
    \label{fig:testdistances}
\end{figure}

Three complementary graph distances are computed using the
\texttt{netrd} \cite{mccabe2020} and \texttt{networkx} \cite{hagberg2008} Python libraries: DeltaCon
\cite{koutra2013}, Ipsen--Mikhailov distance \cite{jurman2011}, and Quantum Jensen--Shannon divergence \cite{nielsen2021}. The
metrics are first evaluated on control sets generated by permuting node
labels of selected reference models while preserving topology. All three
metrics assign lower distances to permuted instances of the same model
than to structurally distinct models, supporting robustness to node
relabeling and sensitivity to structural differences (Figure~\ref{fig:testdistances}).

\begin{figure}[htbp]
\centering
\includegraphics[width=1\textwidth]{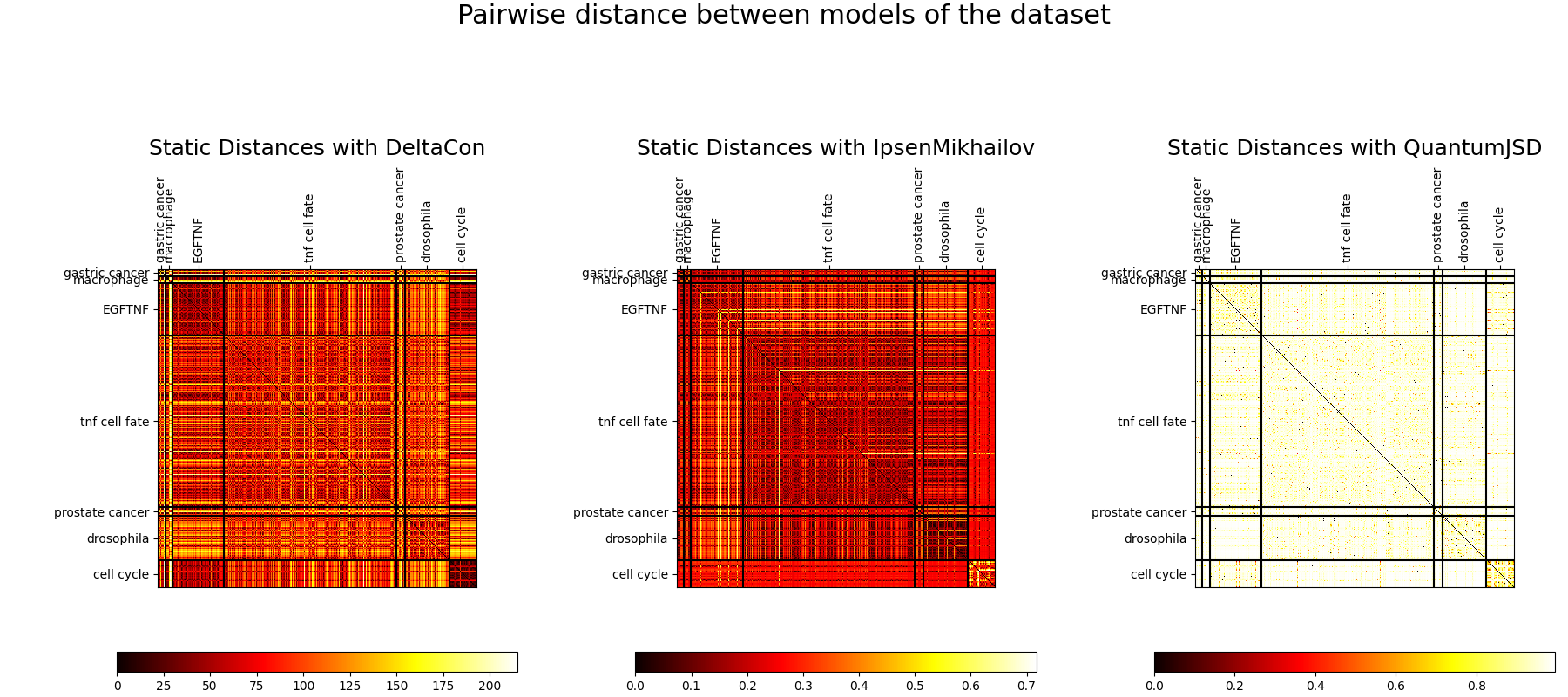}
    \caption{Pairwise structural distances across the curated Boolean model collection. Panels show heatmaps of (a) DeltaCon, (b) Ipsen-Mikhailov distance, and (c) Quantum Jensen-Shannon divergence computed for all filtered models. The matrices summarize structural variability in the released dataset both within and across reference Boolean model families.}
    \label{fig:modeldistances}
\end{figure}

The validated metrics are then applied to the full benchmark suite.
Pairwise distances are computed within and across reference-model
families (Figure~\ref{fig:modeldistances}). Across the three metrics, average within-family
distances are slightly lower than, but comparable to, between-family
distances, with within-to-between distance ratios ranging from 0.76 to
0.85. This indicates that the mutation process generates structurally
diverse models while preserving traceable relationships to their source
architectures.

\subsection{Behavioral heterogeneity of the simulation
dataset}\label{behavioral-heterogeneity-of-the-simulation-dataset}

Behavioral heterogeneity is assessed from multiscale simulation outputs
generated during offline evaluation and model selection. This analysis
tests whether the benchmark models induce diverse responses under
controlled variation in initial conditions and stimulation protocols.

Across selected models, the average correlation between output-response
profiles is 0.06015, indicating low redundancy and largely independent
response profiles. Rare or extreme behaviors are quantified using the
outlier-strength measure used during model selection. Outlier-strength
values span several orders of magnitude, with a maximum of
approximately $1.07 \times 10^{4}$, a mean of 42.44, and quartiles of
0.45, 6.40, and 18.79. Together, these analyses indicate that the
dataset captures both smooth and highly irregular multiscale response
regimes.

\subsection{Scope and limitations}\label{scope-and-limitations}

PhysiBench is an in silico benchmark resource intended for computational
method development and evaluation. It is appropriate for workflows such
as surrogate modeling, parameter estimation, simulation-based
optimization, sensitivity analysis, and algorithmic comparison. Although
the source networks are derived from published biological regulatory
models, the released variants are synthetic constructs generated by
stochastic mutation and filtering. They have not been independently
calibrated or validated against experimental data and should not be used
to infer biological mechanisms, predict cell-line-specific responses, or
draw conclusions about pharmacological efficacy, toxicity, or
therapeutic scheduling in real biological systems.

Several modeling assumptions define the scope of the resource. The
Boolean formalism represents intracellular signaling through discrete
switch-like transitions and does not capture graded signal transduction,
continuous dose-response relationships, or kinetic effects that may be
important in specific biological applications. PhysiBoSS simulations are
executed in a two-dimensional spatial domain, so three-dimensional
tissue architecture, extracellular matrix mechanics, and out-of-plane
diffusion are not represented. The external stimulation protocol is
implemented as an abstract input to the intracellular network and should
not be interpreted as a mechanistic model of drug exposure,
pharmacokinetics, receptor binding, or dose-dependent pharmacological
action.

These limitations define the intended scope of the resource rather than
limiting its use as a controlled benchmark. Users requiring biological
predictive accuracy for a specific system should independently select,
calibrate, and validate an appropriate model before drawing biological
or experimental conclusions.

\section{Usage Notes}\label{usage-notes}

The resource supports three main reuse modes: direct execution of
benchmark models, analysis of the precomputed trajectory dataset, and
regeneration or extension of the dataset using the workflow released in the repository.
Executable examples are maintained in the GitHub repository rather than
embedded in the manuscript, so that commands, paths, and software
requirements remain synchronized with the released code (See Section \ref{codeavail}).

The benchmark model suite can be used directly as a collection of
MaBoSS/PhysiBoSS-compatible intracellular regulatory models. Example
scripts for loading the model manifest, selecting models by reference
family, and launching a PhysiBoSS simulation are provided in the
repository (see Section \ref{codeavail}).
Users may also modify input nodes, output mappings, or simulation
parameters for exploratory computational workflows, provided that
biological interpretation is supported by independent calibration and
validation.

The trajectory dataset can be used for downstream tasks such as
surrogate modeling, data-driven inference, sensitivity analysis, and
simulation-based optimization. Example scripts for reading the
simulation manifest, loading paired input-parameter and cell-data files,
filtering simulations by context parameters, and computing
population-level summaries are provided in the repository (see Section \ref{codeavail})..

The full construction workflow is reproducible through the released
Snakemake pipeline (See Section \ref{codeavail}). The
workflow includes rules for reference-model preparation, mutation-based
generation, online behavioral filtering, offline sensitivity filtering,
model selection, structural validation, and simulation-data extraction.
Users can execute the complete workflow or run individual stages to
regenerate selected components or extend the dataset with additional
simulation contexts. The workflow supports local execution,
containerized environments, and remote execution on high-performance
computing systems through the configuration files provided in the
repository.

\section{Data Availability}\label{data-availability}

The PhysiBench resource, including the benchmark model suite,comprising 612
variant Boolean regulatory network models compatible with PhysiBoSS,  and multiscale trajectory dataset, comprising 120,000 time-resolved PhysiBoSS simulation outputs
generated from 60 selected models under controlled parameterizations, is publicly
available as an open-access record in the Politecnico di Torino
institutional repository \cite{physibench_resource}. The record
is released under an open license and can be reused independently or
together through their shared model identifiers.

\section{Code Availability}\label{codeavail}

All custom code developed to construct, validate, and reuse the resource
is publicly available in the GitHub repository \cite{physibench_repo}. The repository
includes scripts and workflows to reproduce model generation, online
behavioral filtering, offline sensitivity filtering, model selection,
technical validation analyses, simulation-data extraction, and example
downstream reuse workflows. The repository also records the software
versions and execution environment used for dataset generation,
including PhysiBoSS/PhysiCell, MaBoSS, Python dependencies, Snakemake,
and container definitions where applicable. Links to the benchmark-suite
and simulation-dataset records are provided in the repository README.

\end{document}